\documentclass[11pt,a4paper]{JHEP3} 

\usepackage{graphicx}
\usepackage{latexsym}

\newcommand{\bea}{\begin{eqnarray}}
\newcommand{\eea}{\end{eqnarray}}
\newcommand{\be}{\begin{equation}}
\newcommand{\ee}{\end{equation}}
\newcommand{\ph}[1]{\phantom{#1}}

\title{Inflation from N-Forms and its stability}

\author{Tomi S. Koivisto$^{1}$, David F. Mota$^{2,3}$, Cyril Pitrou$^2$\\ $^1$
 Institute for Theoretical Physics, University of Heidelberg, 69120
 Heidelberg,Germany\\ $^2$ Institute of Theoretical Astrophysics, University
 of Oslo, 0315 Oslo, Norway\\ $^3$ Lab. de Physique Theorique et
 Astroparticules, Universite Montpellier II, France\\ e-mail:
 \email{T.Koivisto@Thphys.Uni-Heidelberg.De}\\ e-mail:
 \email{D.F.Mota@astro.uio.no}\\ e-mail: \email{C.X.Pitrou@astro.uio.no}}
\preprint{}	
\abstract{ 
We investigate whether inflation, either
isotropic or anisotropic, may be supported by n-forms.
Canonical field strengths and their duals are taken into account,
 and they are allowed to have a potential and also, when necessary for
 slow-roll, a nonminimal curvature coupling.  New isotropic solutions are
 found for three-forms. 
It is also shown that some n-form actions are equivalent
 to f(R) gravity and scalar field models with possible nonminimal
 couplings. Anisotropic solutions are found for two-forms, generalising vector
 inflation. However, as the later also the two-form is unstable during inflation due to the
 nonminimal coupling to curvature. The stability of the isotropic
 background solutions supported by a triad of vectors is also analysed.}

\begin{document}

\section{Introduction}
%

Inflation is an early era of accelerated expansion of the universe which was
introduced to explain the homogeneity and isotropy of the universe at large
scales \cite{inf1,inf2,inf3,inf4,inf5,inf7,inf8,inf9,inf10,inf11}. Evolution of such a Friedmann-Robertson-Walker (FRW) universe can be
described by a single scale factor.  Whatever the matter contained there, its
energy density behaves like a scalar at large scales, and it is convenient to
consider inflation to be driven by a one or several scalar fields. However, at
a more fundamental level there is no motivation to exclude the possibility of
the energy source of the inflationary expansion having a nonscalar nature. In
particular, higher spin bosonic fields could form condensates, and although
perhaps surprising, slow roll inflation can be made possible even for massive
vector fields, either time-like or space-like
\cite{Koivisto:2008xf,Golovnev:2008cf}. A vector inflation was first proposed
by Ford \cite{Ford:1989me}. Even before, it was known that with the conformal
coupling, the vector field equation of motion is just like that of a scalar
field, however the vector might appearing unstable due to its effective negative
mass-squared during inflation \cite{Turner:1987bw, bertolami}. It is interesting to look
for alternative scenarios to understand how generic such results are.

Such models have been overlooked since they generically induce an
anisotropy. This picture has changed in the last couple of years. The recent
detections of some unexpected features in the CMB temperature anisotropies
have raised a lot of speculations about the need to reconsider some of the
basic cosmological assumptions. A hemispherical asymmetry has been reported
\cite{c4}. The angular correlation spectrum seems to be lacking power at the
largest scales \cite{c3}. The alignment of the quadrupole and octupole (the so
called Axis of Evil \cite{c2}) could also be seem as an extra-ordinary and
unlikely result of statistically isotropic perturbations, even without taking
into account that these multipoles happen also to be aligned to some extent
with the dipole and with the equinox. The Axis of Evil does not show any
correlation with the lack of angular power \cite{c1}.

The significance of the anomalies has been debated extensively in the
literature (see e.g. \cite{f1,f2,f3,f4}) with some reported effects more
significant than others \cite{frode1,frode2,nic}. The difficulty in
quantifying exactly the importance of any effect is due to how correctly the a
posteriori probability of observing is estimated. It is clear however that the
anomalies implying an overall anisotropy in the data are more significant than
the lack of power in the CMB quadrupole. A natural explanation for the
observed anomalies may be some form of a yet undetermined systematic or
foreground signal which is not being taken into account properly in the data
reduction producing the final maps \cite{s1,s2,s3,s4}. However, a conclusive
explanation along these lines has not been put forward yet. It is therefore
legitimate to ask whether the observed anomalies may be an indication of a
departure from the standard cosmological model. CMB data strongly support the
general theory of inflation, we focus here on departures from the standard
picture which can however be reconciled with the inflationary framework.

With these considerations in mind, in this work we perform an initial step
towards the study of a n-form -driven inflation. The basic questions we want to
address are: do we know that the inflaton is a scalar? Is it isotropic?  (By
isotropic we will always mean spatially isotropic).  We mostly consider
models where the comoving field is dynamical, light and slowly
rolling. Comoving field is the square of the form, and as a scalar it is the
suitable degree of freedom to consider. By dynamical we mean that the field
obeys second order equations of motion.  By light we mean that the effective
mass term appearing in the background equation is smaller than the Hubble
rate. By slowly rolling we mean that the derivatives of the field are small
compared to the Hubble rate. All these restrictions are not index necessary
for a successful and viable inflationary scenario: indeed we show that some
models could have only first order equations of motion and that a three-form
can easily inflate the universe even if its comoving field is not slowly
rolling. Our main aim is to find the ${\it simplest}$ possible quadratic
action that is compatible with inflationary solutions in each of the five
classes of models we discuss (in four dimensions, one can have forms with five
different index numbers $0$, $1$, $2$, $3$ or $4$). During the final stages of
this work, other works on form-driven inflation and the generation of
gravitational waves within them have appeared
\cite{Germani:2009iq,Kobayashi:2009hj, germani2,dimopolous}. However, our starting action
(\ref{action}) is different from theirs which includes nonminimal couplings
with coefficients fixed in such a way that the equation of motions are always
exactly of the scalar field field form. Thus there is only partial overlap.
  
The plan of the paper is as follows. In section \ref{Forms} we review general properties of forms. First we motivate the 
basic action (\ref{action}) which we 
write down in subsection \ref{form_intro}, and whose dual we write down in subsection \ref{dual}. 
In subsection \ref{gauge} we consider the St\"uckelberg method of restoring the gauge-invariance of the models by adding a new field, which thus 
illuminates the field content of the theory. In section \ref{cosmo}
we analyse each of the ten cases in turn. We review the vector inflation and show that its two-form generalisation is qualitatively similar in all respects. 
In general, both need nonminimal coupling in the slow-roll case, are not compatible with isotropy and seem to have instabilities. The latter is indicated also by an 
explicit computation. We also analyse the case that the background is made isotropic by considering several mutually orthogonal fields, the so called ''triad'' case. 
In contrast, the three-form does not need nonminimal coupling, is compatible with isotropy and stable. 
In general, the dynamics nonspatial forms is restricted due to the symmetries of the FRW and Bianchi I backgrounds. 
Still, one may construct viable models from them. 
We also find a reformulation of the nonlinear gravity theory (the $f(R)$ models) as a four-form action. 
All these results are concisely summarised in the Table \ref{tab} and in the concluding section \ref{conclu}.

\section{Forms}
\label{Forms}

String theories are generically inhabitated by forms and one could expect such
to appear in effective low energy actions. As they often are accompanied by
axions and ''eat up'' lower order forms, we know that consideration of massive
forms can be well motivated. One possible manifestation of forms is antisymmetric
gravity. Already Einstein attempted a geometric unification of General
Relativity (GR) and electromagnetism by considering an asymmetric
metric. Leading contribution from the antisymmetric part indeed includes a
Maxwell type form, however the theory fails to predict the Lorentz force
law correctly. Nonsymmetric generalisations of GR have continued to be of interest in
theories and also phenomenology \cite{Moffat:1994hv}. In particular, an
antisymmetric tensor field is a crucial ingredient in some gravitational
alternatives for dark matter \cite{Prokopec:2006kr}. On the other hand, this
presents a natural way to generate propagating torsion. General actions for
forms can break gauge invariance which easily could awake ghost or tachyon
modes that otherwise remained unphysical gauge modes. Nonsymmetric theories
may thus be stringently constrained, though at nonlinear level some
instabilities could be avoided \cite{Damour:1992bt,Clayton:1996dz}. However,
some simple stable actions exist.

Motivated by these theoretical interests in unified theories at both classical
generalisations of GR and in fundamental theories for quantum gravity, in the
present study we consider the possibility to employ the stable types of the
form field actions in cosmology where unknown fields are indeed needed, namely
as energy sources for the present acceleration and inflation. The quantum
generation of two-form fields during inflation has been considered
\cite{Prokopec:2005fb}. Here we instead study form field driven inflation.
Presence of n-forms could also explain the origin of four large dimensions,
since due to inherent anisotropy of forms one might consider scenarios where
only three spatial directions inflate while extra dimensions were stabilised
\cite{ArmendarizPicon:2003qw}. However, our focus here is on a possible
residual anisotropy that could be (or even could have been as discussed in the
introduction) observed in CMB, and in cosmological calculations (from the next
section onwards) take a four dimensional action as our starting point.

To motivate the particular form of the action we consider, we begin by
reviewing some general results about two-forms. In flat space, the most
general quadratic, Lorentz- and parity invariant Lagrangian reads for a
two-form $A$ as
\be \label{lag}
\mathcal{L}_f = -\frac{1}{4}a(\partial A)^2 - \frac{1}{2}b(\partial \cdot A)^2 - \frac{1}{4}m^2 A^2,
\ee
where the dot product means contraction over the first indices. van
Nieuwenhuizen has shown that unless $a(a+b)=0$ the propagator contains
nonlocality or a ghost \cite{VanNieuwenhuizen:1973fi}. This leaves two
possibilities. If $a=-b$, one may show by partial integrations that the
Lagrangian reduces to $\mathcal{L}_f = -aF^2/48 - m^2A^2/4$, where $F$ is the
Maxwell tensor formed from $A$, which in general is defined for a $n$-form as
follows
\be \label{maxw}
F_{M_1\dots M_{n+1}}(A) = (n+1)\partial_{[M_1}A_{M_2\dots M_{n+1}]}\,,
\ee
where the square brackets indicate antisymmetrisation, for example $A_{[M N]}=
(A_{MN}-A_{NM})/2$ and so on.  The other possibility, that $a=0$, corresponds
to the dual theory, as will become clear in \ref{dual}. Thus, in the flat
space limit our theory ought to reduce to either a massive Maxwell i.e. Proca
theory, or its dual.  (One knows that this applies to the vector case as well
since then, in addition to the Maxwell term, a gauge-fixing term which is got
by $a=0$ and actually reduces to a scalar theory, is the only consistent
choice for a kinetic term in vacuum).

Apparently, in curved space more possibilities emerge. The form field might
have couplings with the curvature tensors. The general quadratic couplings
then include contributions from the three terms
\be \label{lagc}
 \mathcal{L}_c = -\frac{1}{2}\sqrt{-g}\left(\xi R A^2 - c R_{MN}A^{N K}A_K^{\ph{K}M} - d 
R_{MNKL}A^{MN}A^{KL}\right)\,.
\ee
Note that for example couplings of the type $RF^2$,
$R_{MN}F^{MK}F_K^{\ph{K}N}$ or $R_{MNKL}F^{MN}F^{KL}$ would result in
higher-order derivative theory. Thus we leave them out, though it is well
known such could be motivated by quantum corrections \cite{Drummond:1979pp}
(for recent higher order gravity-vector investigations see
\cite{Bamba:2008xa,Bamba:2008ja}).  However without unreasonable fine tunings,
the stability of cosmology requires $c=d$, and to have always stable
Schwarzchild solutions one must further set $d=0$ \cite{Janssen:2006nn}. Hence
we are left with only a possible nonminimal coupling to the Ricci scalar
$R$. Therefore our starting point will be a Maxwell action with a mass and a
curvature coupling terms as free parameters. The mass can be promoted to a
more general potential function $V(A^2)$ without complicating the analysis.

\subsection{Equations of motion}
\label{form_intro}

Thus, we consider an n-form $A$ in $d$ dimensions with the following action
\be \label{action}
S = \int d^dx\sqrt{-g}\left(\frac{1}{2\kappa^2}R - \frac{1}{2(n+1)!}F^2 - V(A^2) - \frac{1}{2 n!}\xi A^2 R\right),
\ee
where $\kappa = 1/\sqrt{8\pi G_N}$ where $G_N$ is the Newton's constant. The kinetic term is given by the square of the field strength,
\bea 
\label{strenght}
F^2 = F_{M_{n+1}}F^{M_{n+1}}, \qquad F \equiv (n+1)\left[\partial A\right]\equiv {\rm d} A\,  
\eea
In the cases where the indices are not written explicitly, the big square
brackets mean antisymmetrisation. The last equality also defines the exterior
derivative.  Squaring means, here and elsewhere, contracting the indices in
the same order. In the following, dotting means contracting the first index,
and gradient means adding an index by differentiating.  Note that the second
line (\ref{strenght}) is nothing but the definition (\ref{maxw}) in the new
more compact notation. We also use the abbreviation
\be
M_n \equiv M_1\dots M_{n}\,.
\ee
With these notations, the stress tensor found by metric variation can be written as
\bea \label{stress}
T_{MN} & = & \frac{1}{n!}F_{M M_{n}}F_N^{\ph{N}M_n} + 2 n V'(A^2)A_{M M_{n-1}}A_N^{\ph{N}M_{n-1}}
-g_{MN}\left(\frac{1}{2(n+1)!}F^2 + V(A^2)\right) \nonumber \\
& + & \frac{\xi}{n!}\left[n R A_{M M_{n-1}}A_N^{\ph{N}M_{n-1}} + \left(G_{M N} - \nabla_M\nabla_N + g_{MN}\Box\right)A^2 \right],
\eea
where the second line is the contribution from nonminimal coupling. We note that the kinetic piece is traceless 
when $d=2(n+1)$: the Maxwell field in $d=4$ is conformal
i.e. traceless. Furthermore, the potential piece is 
traceless if $V(x)\sim x^p$ where $p=d/(2n)$: a $(d/2-1)$-form in any $d$
could be conformal with an interaction 
term $p=d/(d-2)$. 

The equations of motion are
\be \label{eom}
\nabla \cdot F = \left(n! 2 V' + \xi R\right)A,
\ee
implying thanks to antisymmetry
\be \label{eom2}
\nabla \cdot \left[\left(n! 2V' + \xi R\right)A\right] = 0\,.
\ee

\subsection{The dual theory}
\label{dual}

There are various theories of fundamental physics where duality transformation plays a role. 
Therefore it is motivated to consider also the dual actions for each
$n$-form. If the forms are nonminimally coupled 
to gravity, the usual duality 
invariance does {\it not} hold at the level we consider. 
The Hodge duality transforms an $n$-form into a $(d-n)$-form. Component wise, the transformation rule is
\be \label{cwise}
(\ast A)_{M_{d-n}} = \frac{1}{n!}\epsilon_{N_{n}M_{d-n}}A^{N_{n}}. 
\ee
It follows that $\ast(\ast A) = {\rm sgn}(g)(-1)^{(d-n)n}A$.
The strength of a dual is not the same as the dual of the strength. The Hodge
star operator should be applied to each 
form, namely to the strength 
appearing in the action (\ref{action}), as it is an $(n+1)$-form. We get
\bea 
(\ast F)_{M_{d-n-1}} &=& \frac{1}{(n+1)!}\epsilon_{N_{n+1}M_{d-n-1}}F^{N_{n+1}}.
\eea
From this follows, by using the identity (\ref{cwise}) and the definition (\ref{strenght}), that
\be
(\ast F) = (-1)^{n}\nabla \cdot (\ast A).
\ee
Note also that by defining $\delta \equiv {\rm sgn}(g)(-1)^{dn+d+1}{\ast} {\rm
  d} {\ast}$, this can be recast in the more 
compact form 
$(\ast F) = (-1)^{n+1}\delta (\ast A)$. For a review of differential geometry and gauge theories see \cite{Eguchi:1980jx}.
We find also $(\ast A)^2 = {\rm sgn}(g)(d-n)!A^2/(n!)$. Thus a spacelike field
is transformed into a timelike and vice 
versa. The dual action for (\ref{action}) 
is then written as
\bea \label{actiond}
S = \int d^dx\sqrt{-g}\left[\frac{1}{2\kappa^2}R -
  \frac{1}{2(d-n-1)!}\left(\nabla\cdot (\ast A)\right)^2 
  -  V\left((\ast
  A)^2\right) - \frac{1}{2(d-n)!}\xi(\ast A)^2 R \right].
\eea
Thus the Maxwell-type kinetic term transforms to a square of a divergence. We
see that indeed the dual has the unique 
form that is 
stable in flat space, got by setting $a=0$ in Eq.(\ref{lag}).
The contribution to the stress tensor from this term constitutes the following
two lines, the third is due to the 
potential,
and the fourth vanishes if coupling to gravity is minimal:
\bea \label{stressd}
\ast T_{MN} & = & \frac{1}{(d-n-1)!}\Big[2\left(\nabla_{(M}(\ast A)_{N) M_{d-n-1}}\right)\left(\nabla_K(\ast A)^{K 
M_{d-n-1}}\right)  \\ 
& + & (d-n-1)\left(\nabla_L(\ast A)^L_{\ph{L}M M_{d-n-2}}\right)\left(\nabla_K(\ast A)^{K\ph{N}M_{d-n-2}}_{\ph{K}N}\right) - \frac{1}{2}g_{M N}\left(\nabla\cdot 
(\ast A)\right)^2\Big] \nonumber \\ 
& + & 2(d-n) (\ast A)_{M M_{d-n-1}}(\ast A)_N^{\ph{N}M_{d-n-1}}V'((\ast A)^2) - g_{MN}V((\ast A)^2)  \nonumber\\
& + & \frac{\xi}{(d-n)!}\left[(d-n) R (\ast A)_{M M_{d-n-1}}(\ast A)_N^{\ph{N}M_{d-n-1}} + \left(G_{M N} - \nabla_M\nabla_N + g_{MN}\Box\right)(\ast 
A)^2\right].\nonumber
\eea

The potential and coupling terms are similar to those in (\ref{stress}). We
note a massless dual is traceless if 
$n=d/2+1$, 
and a power-law potential leaves no trace if $p=d/(2(d-n))$.  
The equations of motion for the field are
\bea \label{eomd}
\left[\nabla \nabla \cdot (\ast A)\right] = 2\left((d-n-1)!V'+\frac{1}{2(d-n)}\xi R\right)(\ast A).
\eea
Note that we can deduce only about the antisymmetric part of the variation since we are varying with respect to $(\ast A)$.
One can show that the $(d-n)$-form with an action (\ref{actiond}) can be equivalent to a canonical $(d-n-1)$-form $\phi_{M_{d-n-1}}$ by introducing 
Lagrange multipliers to fix $\phi = -\nabla \cdot (\ast A)$ The resulting action is 
\be \label{actiond2}
S = \int d^dx\sqrt{-g}\left[\frac{1}{2\kappa^2}R - V((\ast
  A)^2((\nabla\phi)^2)) -  \frac{1}{2(d-n-1)!}\phi^2 - \frac{1}{2(d-n)!}\xi (\ast A)^2((\nabla\phi)^2)R 
\right],
\ee
where $(\ast A)^2$ is given by inverting 
\be \label{cons}
\left(\frac{\xi}{(d-n)} R + 2(d-n-1)!V'((\ast A)^2)\right)^2 (\ast A)^2 = \left[\nabla\phi\right]^2.
\ee
If $V((\ast A)^2) = \frac{1}{2 (d-n)!}m^2(\ast A)^2$ and we rescale $\phi \rightarrow \phi m$, we get
\bea \label{actiond3}
S = \frac{1}{2}\int d^dx\sqrt{-g}\left[\frac{1}{\kappa^2}R - \frac{1}{(d-n)!(\xi R/m^2 + 1)}[(d-n) \nabla\phi]^2 
\nonumber  - \frac{m^2}{(d-n-1)!}\phi^2\right].
\eea
If $\xi=0$, this is nothing but the canonic $(d-n-1)$ form since $[(d-n)
  \nabla \phi]={\rm d} \phi$ is the strength 
of $\phi$.
In that case, the duality becomes trivially transparent in a more elegant notation \cite{Eguchi:1980jx}. Since the 
equation of motion for the dual can be written as
\be
{\rm d} \delta(\ast A) + m^2(\ast A) = 0,
\ee 
and if we choose $m \phi = \delta(\ast A)$, it is clear that the two terms
that appear in the quadratic action can
 be expressed in terms of $\phi$ as
\bea
(\ast F)\wedge(\ast\ast F) & = & m^2 \phi\wedge(\ast\phi), \\
m^2(\ast A)\wedge(\ast\ast A) & = & {\rm d}\phi\wedge(\ast{\rm d}\phi),
\eea
where by $(\ast F)$ we mean of course $(-1)^{n+1}\delta(\ast A)$.
Let us make some remarks about different cases. We note that to get a more
general potential for the field $\phi$, one would need to consider some
general function of the dual kinetic term, $f(x)$, where $x=(\nabla\cdot(\ast
A))^2$. Such nonlinear models are outside the scope of the present study, but
in Table \ref{dualtab} we summarise briefly the relations between the
equivalent formulation in more general cases than shown explicitly in here. On
the other hand, if the field $(\ast A)$ has a more general potential than a
mass term, the resulting reformulation naturally exhibits noncanonical kinetic
terms. In the case $n=d-1$, it is thus a way to generate k-essence models
\cite{ArmendarizPicon:1999rj} as pointed out by Gruzinov
\cite{Gruzinov:2004rq}. 

If $\xi\neq 0$ we obtain an unusual coupling between the generalised scalar
field and curvature; previously such couplings have been applied in attempts
to model a dynamical relaxation of the cosmological constant, dark matter and
gravity assisted dark energy
\cite{Dolgov:2003fw,Nojiri:2004bi,Koivisto:2005yk,Sotiriou:2008it}. 

\TABLE{
\begin{tabular}{|c|c|c|}
\hline
$(\ast A)$               & $\phi$ &  $n=d-1$ \\
\hline
Mass term          &  Canonical kinetic &  Quintessence \\
\hline
General potential  &  Noncanonical kinetic &  K-essence \\
\hline
\hline
Dual kinetic        &  Mass term         &  Chaotic inflation \\
\hline
Function  &  General potential &  General scalar \\
\hline
\end{tabular}
\caption{\label{dualtab} The correspondence between an original dual $(d-n)$ form $(\ast A)$ and the $(d-n-1)$ form field $\phi$ in the
reformulation of the theory. The canonic dual with a mass can be rewritten as a massive gradient kinetic term, but more general Lagrangians are 
obtainable. In the last column we have indicated the type of a corresponding scalar field. The potential of the quintessence is given by the form of the
kinetic term of the field $A$: in case of a canonical kinetic term of $A$, the quintessence model has a canonical mass term. This is also what we mean by 
the last line: a given function of the dual kinetic term , turns into a given potential function of the quintessence field.
}
}


\subsection{Gauge invariance}
\label{gauge}

The potential terms manifestly break gauge invariance. By performing the St\"uckelberg trick, we can promote the
action into a gauge invariant form. Then we introduce a new field patterned after the gauge symmetry, whose transformation compensates
for the gauge symmetry violation. Only the antisymmetric part of the field counts, so in essence we will have an $(n-1)$-form as the following:
\be
A = B + \frac{1}{m}\left[\partial \Sigma\right],
\ee
where $m>0$ is a suitable mass scale. The new Lagrangian density,
\bea
\mathcal{L} = -\frac{1}{2(n+1)!}F^2(B) - \frac{1}{2}M^2\left(B + \frac{1}{m}\left[\partial \Sigma\right]\right)^2,  
\eea
where $M^2$ is given by the effective mass including the contribution from the coupling, then has the gauge symmetry
\bea
\Sigma  \rightarrow \Sigma + \Delta, \qquad \qquad
B         \rightarrow  B - \frac{1}{m}\left[\partial\Delta\right],
\eea
where $\Delta$ is an arbitrary tensor with $(n-1)$ indices.
In the gauge $\Sigma = 0$ we recover our original Lagrangian. We see that the field $\Sigma$ has a kinetic term,
and also a cross term. If we choose the specific gauge where the gradient of the $\Sigma$ is orthogonal to
our $n$-form\footnote{Alternatively one could reach our result by considering the limit $M^2 \rightarrow 0$. This 
limit is smooth and preserves the degrees of freedom, and the amount, though not the form of the gauge 
symmetry.} and set $m=\sqrt{|M^2|}$, the Lagrangian density becomes 
\bea \label{lagr}
\mathcal{L} = -\frac{1}{2(n+1)!}F^2(B) - \frac{1}{2}M^2 B^2 - 
\frac{1}{2}{\rm sgn}(M^2)\left[\partial\Sigma\right]^2.  
\eea
Now the $\Sigma$ field, or more precisely the $(n-1)$-form one gets by anti-symmetrising the $\Sigma$, 
is a ghost if $M^2$ is negative. We have reached the conclusion that if the effective mass of an $n$-form is 
negative, there is an $(n-1)$-form ghost degree of freedom in the theory.

\section{Homogeneous Cosmology}
\label{cosmo}

In this part we consider the use of $n$-forms for early cosmology and we thus restrict to the case $d=4$ and to homogeneous cosmologies (that is the fields depend only on time). We consider the most simple metric compatible with anisotropy when required, 
namely an axisymmetric Bianchi I metric, which can be written as
\be 
\label{metric}
ds^2 = -dt^2 + e^{2\alpha(t)}\left[e^{-4\sigma(t)}dx^2 + e^{2\sigma(t)}\left(dy^2+dz^2\right)\right].
\ee
We also define $H=\dot{\alpha}$ and we will refer to $\dot \sigma$ as the shear. 
The reason we restrict ourselves to axisymmetry is that in a general Bianchi I spacetime the stress tensor (\ref{stress}) is 
compatible with only one nonzero component of any $n$-form, and this results in an axisymmetric stress energy tensor. A caveat
is that one may allow, in some cases, more components if they satisfy certain first order constraint equations. 
Fields obeying such constrains would not be dynamical in the usual sense, and typically exhibit decaying anisotropic stress. 
We do not consider these special cases in more detail.

\subsection{Zero-form}
\label{zero}
   
A zero-form is a scalar field and its application to inflationary cosmology has been intensively studied in the literature over the three past 
decades~\cite{Brans:1961sx,Wetterich:1987fm,muk,cop}. Since a scalar field has a vanishing contribution to the 
anisotropic stress tensor, the anisotropy is decaying ($\dot \sigma \sim 1/a^3$) and one often study the cosmologies with scalar fields directly in a homogeneous and isotropic universe, that is with FRW symmetries.


\subsection{One-form}
\label{one}

Only one nonzero component is allowed by the Bianchi I symmetry, having two components would introduce nondiagonal components in the stress tensor 
(\ref{stress}). 
A zero component would be nondynamical, its equation of motion simply stating it vanishes or that the $V' \sim R/2$. 

So let's consider spatial component $$A_M=e^{\alpha(t)-2\sigma(t)}X(t)\delta_{M x}.$$ 
Then $A^2(t)=X^2(t)$. The equation of motion (\ref{eom}) becomes
\bea \label{eom_a}
\ddot{X}+3H\dot{X}+\left[2V'+(1+6\xi)(\dot{H}+2H^2)- 2H\dot{\sigma}-2\ddot{\sigma} - (4-6\xi)\dot{\sigma}^2\right]X = 0,
\eea
Now the coupling $\xi=-1/6$ eliminates the effective mass terms. 
However, the model is unstable, as we see in subsection \ref{p_1} and was first discussed in detail in \cite{Himmetoglu:2008hx}.
Since the effective mass squared is negative, there is a scalar ghost, in accord with our generalised argument in 
subsection \ref{gauge}.

Now if we want to understand if this slow-roll dynamics is compatible with
inflation with small anisotropy we need $\rho,P$ and $\Pi$ where the last quantity
is defined from the anisotropic stress tensor 
$\pi^i_{\,j}=$diag$(-2\Pi,\Pi,\Pi)$. We specialise here to $V(x)=\frac12 m^2 x$.
By the usual definitions, the energy density, isotropic and anisotropic pressure come out as
\begin{eqnarray}
\rho=\frac{1}{2}\left[m^2 X^2 +(\dot X+ HX -2 \dot \sigma X)^2 \right] 
+ \xi \left[ 6 H X \dot X +3 (H^2-\dot \sigma^2)X^2 \right]\,,
\end{eqnarray}
\begin{eqnarray}
&&P=\left[\frac{1}{6}(\dot X+HX-2 \dot \sigma X)^2-\frac{1}{6}m^2 X^2\right]\\
&&-2 \xi \left[\dot X^2 -H X \dot X- X^2\left( m^2+(1+6 \xi) \dot H +H^2 \left(\frac{5}{2}+12\xi\right)-2 H \dot \sigma -2 \ddot \sigma - \left(\frac{9}{2}-6 \xi\right)\dot \sigma^2 \right)\right]\nonumber\,,
\end{eqnarray}
\begin{equation}\label{pi_vec}
\Pi= \frac{1}{3}\left[(\dot X+HX-2 \dot \sigma X)^2-m^2 X^2\right] - \xi \left[2\dot H + 4 H^2 +2 \dot \sigma^2 -(\ddot \sigma + 3 H \dot \sigma) \right]X^2+\xi 2 \dot \sigma X \dot X\,.
\end{equation}
The general condition for the violation of the strong energy condition (and thus acceleration) is $\rho+3P< 0$.
In the minimally coupled case ($\xi=0$), we see directly that 
\begin{equation}
\rho+3P = (\dot X+HX-2 \dot \sigma X)^2 > 0\,,
\end{equation}
and there is no possible accelerated expansion, that is no inflation.

However in the case $\xi=-1/6$ which is required to ensure slow-roll of the field
\begin{equation}
\rho + 3 P = 2 \dot X^2 -4 \dot \sigma X \dot X + X^2 (10 \dot \sigma^2 - m^2 + 2 \ddot \sigma -2 \dot \sigma H)\simeq \left(2 \ddot \sigma-m^2\right) X^2\,,
\end{equation}
from which the equation for $\ddot a/a$ can be deduced thanks to 
\begin{equation}
\frac{\ddot a}{a}= -\frac{\kappa^2 (\rho+3 P)}{6}-2 \dot \sigma^2\,.
\end{equation}
If we can neglect $2 \ddot \sigma$ in front of $m^2$ and assume a negligible shear $\dot \sigma \ll H$, then this leads apparently to 
exponential inflation [$a\sim \exp(Ht)=\exp(m\kappa Xt/\sqrt{6})$]. 
However, it turns out that we can define an effective anisotropic stress is given by
\begin{eqnarray}
\tilde \Pi\equiv \Pi+\frac{1}{6}(\ddot \sigma + 3 H \dot \sigma)X^2\nonumber
=\frac{1}{3}\left[ \dot X^2 +X \dot X (2H - 5 \dot \sigma)+ X^2 (-4 \dot \sigma H +5 \dot \sigma^2 - m^2 + \dot H + 3 H^2 )\right]\,,
\end{eqnarray}
in such a way that the evolution equation for the shear can be written as
\begin{equation}
\left(1+\frac{\kappa^2}{6}X^2\right)\left(\ddot \sigma+3 H \dot \sigma\right) =\kappa^2 \tilde \Pi\,.
\end{equation}
The dominant term on the right hand side of this equation is $\kappa^2 H^2 X^2$ and tends to increase the shear.
For a field large enough ($\kappa^2 X^2/6 \gg 1$) which is anyway required to have enough $e$-folds of accelerated expansion, first $\ddot \sigma \simeq 6H^2$ with $\dot \sigma \simeq 0$,  and $2\ddot \sigma > m^2$ from which we obtain $\rho+3P<0$. 
Additionally, after a transition period the shear is reaching $\dot \sigma \simeq 2 H$ with $\ddot \sigma \simeq 0$. We deduce that these two conclusions contradict exponential inflation and negligible shear. 
Thus there is no inflation, that is accelerated expansion, with just one slow-rolling vector field.\\

\subsection{Triads}

Consequently, we need to assume that there are not one but several vector fields. The simplest model assumes a triad of vector fields~\cite{Golovnev:2008cf}, to ensure that there is no shear ($\dot \sigma=0$). If we do so, we can assess quickly the 
fine-tuning involved. We can parameterise the difference of one vector field with respect to the two other vector fields by
\begin{equation}
X_1^2=(1+\epsilon)X^2/3, \quad X^2_2=X^2_3\equiv X^2/3 
\end{equation}
with $\epsilon \ll 1$.
Then if the fields are slow-rolling, if there is no shear to start with ($\dot X_i \ll H X_i$ and $\dot \sigma \ll H$) and if the field is rather large in order to ensure long enough inflation ($\kappa^2 X^2 /6 \gg 1$), we obtain as long as the shear 
stays negligible 
\begin{equation}
H^2\simeq\frac{\kappa^2}{6}m^2 X^2\,,
\end{equation}
\begin{equation}
\rho+3P \simeq X^2 \left[- m^2+ 2 \epsilon  \ddot \sigma \right]\,,
\end{equation}
\begin{equation}
\tilde \Pi \simeq \epsilon X^2 H^2,\qquad  \ddot \sigma \simeq 6 \epsilon H^2  \,.
\end{equation}
In order to have accelerated expansion, we thus need 
\begin{equation}
\epsilon < m^2/(2 \ddot \sigma)\,,
\end{equation}
which leads to
\begin{equation}
\epsilon < (\sqrt{2} \kappa X)^{-1}.
\end{equation}
Since the number of $e$-folds is approximately given by
\begin{equation}
N \simeq \left(\frac{\kappa X}{2}\right)^2-\frac{1}{2}\,,
\end{equation}
then in order to obtain the $N \ge 70$ $e$-folds required to solve the problems of the standard hot big-bang model, we need $\kappa X > 17$ and thus the condition on $\epsilon$ is 
\begin{equation}
\epsilon < 0.04\,\, .
\end{equation}
The model with a triad of vector fields is thus fine-tuned but the fine-tuning is only going like $1/\sqrt{N}$ and can be considered as reasonable.

\subsection{Two-form}
\label{two}

If we take 
\be \label{twot}
A_{M N} = e^{\alpha(t)-2\sigma(t)}X(t)(\delta_{M0}\delta_{Nx}-\delta_{Mx}\delta_{N0}),
\ee
then $A^2(t) = -2X^2(t)$. However, now the kinetic term is identically zero, and the equation of motion (\ref{eom}) dictates an algebraic 
constraint for the field $X(t)(4V'+\xi R)=0$, analogous to the vector case but
now living in an anisotropic background.  Such constrained models could be
applied to construct effectively nondynamical cosmological fields like in the
so called Cuscuton models or in modified gravity within the Palatini approach
\cite{cus,cus2,Koivisto:2005yc,Koivisto:2007sq}. Now however, the underlying
theory does have more degrees of freedom. Here the fields are restricted only
due to the homogeneity of the FRW or Bianchi I background, and therefore the
perturbations about this background can propagate. Thus in principle they
could also be responsible for the primordial spectrum of perturbations and
structure in the universe. Furthermore, this means also that these models are
not trivialised at large scales by averaging, which might occur for gravity
modifications of the Palatini type \cite{pal1,pal2}.  The formal reason for
the time-like fields being algebraically constrained is that that the Bianchi
I symmetry allows only time derivatives, but the zero index acting in the
kinetic term must vanish in its antisymmetrisation if it appears in a
component of the field.

Here we however consider a space-like two-form. Therefore let us take
\be \label{twos}
A_{M N} = e^{2\alpha(t)+2\sigma(t)}X(t)(\delta_{My}\delta_{Nz}-\delta_{Mz}\delta_{Ny}). 
\ee
Then $A^2(t)=2X^2(t)$. The constraint (\ref{eom2}) is identically satisfied by this ansatz. The equation of motion (\ref{eom}) yields 
\bea
\ddot{X} &+& 3H\dot{X} + 2\left[2V'(A^2) + (1+3\xi)\dot{H} \right. \\ &+& \left. (1+6\xi)H^2  
- \dot{\sigma}H + \ddot{\sigma} - (2-3\xi)\dot{\sigma}^2 \right]X = 0.  \nonumber
\eea
The coupling $\xi=-1/6$ allows now to eliminate the effective mass due to $H^2$. Then a slow-roll suppressed mass term
remains due to $\dot{H}$, plus the shear-terms. These are small and don't (necessarily) spoil slow-roll.  
However, since the effective mass squared is negative, there is a vector ghost, which follows directly from our 
generalised argument in subsection \ref{gauge}. This can also be confirmed explicitly by considering the perturbations in 
Minkowski space, see subsection \ref{p_2}.

Again defining the anisotropic stress as $\pi^i_{\,j}=$diag$(-2\Pi,\Pi,\Pi)$, we find now that
\bea
\Pi & = & + 6 H^2 + 4\dot{H} - 2V + 4V'X^2
+ 6\sigma^2 + \ddot{\sigma} + 3H\sigma \\ \nonumber  
& + & \xi\left( - 4\ddot{X}X - 4\dot{X}^2 + 2X^2\dot{H} + 3H^2X^2 - X^2\ddot{\sigma} + \dot{X}X\dot{\sigma} - 8 \dot{X}XH - 3X^2\dot{H}\dot{\sigma}\right).
\eea
This is similar to with the vector expression (\ref{pi_vec}) of the vector case. Thus, 
following similar arguments as there, one could deduce that a single two-form cannot support an 
inflating background for many $e$-folds, since the anisotropy of the solution tends to grow quickly. 
Further, one could again add a triad of two-forms to ensure an isotropic background. 
These three two-forms would again have to be tuned to be equal with the (qualitatively) same accuracy as in the vector case.  
The similarity of the vector and two-form cases can be traced to the duality discussed in general terms in section \ref{dual}. In the next section we apply 
it in more detail in the specific cases of vector $\leftarrow$ two-form and two-form $\leftarrow$ vector (these cases are not equal due to the nonminimal 
coupling).  

%
%

\subsection{Three-form}
\label{three}

If we would consider the ansatz $A = e^{2\alpha+\sigma}X(t)dt\wedge dy \wedge dz$, we would again have an algebraic model,
$X(12V'-\xi R)=0$. Thus, like in the previous cases, we consider only spatial indices. 
Since there are three of them, no direction is picked up, and we
can restrict to the case of FRW, $\sigma(t)=0$. We write $A=e^{3\alpha(t)}X(t)dx\wedge dy\wedge dz$. It follows that $A^2(t)=6X^2(t)$.
The Friedmann equation we obtain is
\bea
3\left[\frac{1}{\kappa^2}-\left(\frac{3}{2}+\xi\right)X^2\right]H^2 =
\frac{1}{2}\dot{X}^2 + 3(1+2\xi)H\dot{X}X + \tilde V(X^2) 
\eea
and the equation of motion for the field is
\be \label{3eom}
\ddot{X} + 3H\dot{X} + 3\left[\frac{2}{3}\tilde V'(X^2) + 4\xi H^2 + \left(1+2\xi\right)\dot{H}\right]X = 0\,,
\ee
where $\tilde V(X^2) \equiv V(A^2)= V(6X^2)$. Thus, the nonminimal coupling introduces an extra mass term, just like for a scalar 
field. We set $\xi=0$ in the following. 
Thus we consider a different case from Refs.\cite{Germani:2009iq,Kobayashi:2009hj}, where suitably fixed  
nonminimal gravity couplings were used to turn the equation of motion (\ref{3eom}) into the Klein-Gordon form.
Now the effective energy density and pressure may be then written as
\bea
\rho_X & = & \frac{1}{2}\left(\dot{X}+3HX\right)^2 + \tilde V(X^2),  \\  
P_X    & = & -\frac{1}{2}\left(\dot{X}+3HX\right)^2 + 2 \tilde V'(X^2)X^2-\tilde V(X^2).
\eea
With quadratic potential the behaviour is the {\it reverse} of that of a scalar field with Hubble friction contributing to the kinetic energy:
the kinetic piece gives a negative, the potential a positive contribution to the pressure. If $V'$ can be neglected for some general potential 
though, both contributions are negative. Then the equation of state mimics a cosmological constant though the field could evolve. 
The condition for inflation is that
\be
\frac{6\ddot{a}}{\kappa^2 a} = \left(\dot{X}+3HX\right)^2 + 2\tilde V(X^2)-6 \tilde V'(X^2)X^2 > 0.
\ee 
Thus, we do not need slow roll to get inflation. A three-form seems to accelerate a FRW universe more naturally than a scalar field. 
To realise phantom inflation in this model one needs a negative slope for the potential,
\be
\dot{H} = -\kappa^2 X^2 \tilde V'(X^2) > 0.
\ee
Now the model seems stable, since the St\"uckelberg argument does not give a ghost when $V>0$, and since we have just a canonical action without the dangerous nonminimal couplings. The quantitative predictions of this class of models will be considered elsewhere. 

%

\subsection{Four-form}
\label{four}

The kinetic term for a four-form is trivial in four dimensions. The fact that such term can lead to a constant contribution to energy density has been
employed in an anthropic solution to the cosmological constant problem \cite{Turok:1998he}. With general potential however, the field can have nontrivial
contribution. The variation of the action with respect to the field $A^2=\varphi$ leads
to an algebraic constraint for the field, $48V'(\varphi)=\xi R$. If the solution is plugged back into the action, we recover a higher order 
gravity theory in the form of a metric $f(R)$ theory
\be \label{actionfr}
S = \int d^4 x\sqrt{-g}\left[\frac{1}{2}\left(\frac{1}{\kappa^2} - \xi\varphi(R)\right)R - V(\varphi(R))\right].
\ee
When $\xi=0$ this reduces to general relativity with a cosmological constant
given by the minimum of $V$. If $V$ is a mass term for the potential $A$, only
constant-Ricci solutions are compatible with such a theory. Thus our world is
not compatible with a simple $V(x) \sim x$ with coupling to the curvature.

One more interesting simple example is a self-interaction which may be seen as
a mass term for the field squared, $V(x) \sim x^2$. This results in
curvature-squared $R^2$ correction to the Einstein-Hilbert action.
Furthermore, one notes that there exists also a trivial solution to the
equation of motion, $A=0$. Thus there are two branches of, GR and
$f(R)$. Perhaps the other branch could be used to ''switch off'' the $f(R)$
theory like in some kind of phase transition when $A$ crosses zero. On the
other hand, to exclude the GR solutions, one could add a $1/A^2$ term in the
potential.

%
\TABLE{
\begin{tabular}{|c|c|c|c|c|c|c|}
\hline
 $n$        & \# dof &  $\Pi$       & $\xi$     &  Sec.              & Comment                         &  Ref.                           \\
\hline
\hline
$0$         & $ 1 $  &  0             & 0              & \ref{zero}    & A scalar field                     &   \cite{Linde:1981mu}        \\
\hline
\hline
$1$         & $ 4 $  & $\sim X^2$      & $-\frac{1}{6}$ & \ref{one}     & Needs nonminimal coupling for SR   & \cite{Golovnev:2008cf}       \\
\hline
\hline
$2$ & $ 6 $  &  $\sim X^2$    & $-\frac{1}{6}$        & \ref{two}     & Needs nonminimal coupling for SR     & -                \\
\hline
\hline
$3$ & $ 4 $  &  0                    &    0            & \ref{three}   & isotropic SR inflation             &   -                           \\
\hline
\hline
$4$ &        $ 1 $  &  0             &  $\neq 0$       & \ref{four}    & metric $f(R)$ gravity              &  \cite{Starobinsky:1980te}     \\
\hline
\end{tabular}
\caption{\label{tab} A summary. For each class of forms, we indicate 1) the number of degrees of freedom (in the cosmological background, the 
effective number could of course be less) 2) the anisotropic stress of the model (which is zero for isotropic cases) 3) the section in this paper 
where we focus on the case 4) a comment about the general nature of model 5) a reference for some related earlier study.}
}

\section{Perturbations}

\subsection{0-form}

This is a scalar field, and the general Bianchi I case (not necessarily axisymmetric) perturbation theory is built in \cite{Pereira:2007yy} and slow roll inflation is studied in \cite{Pitrou:2008gk}. The axisymmetric Bianchi I case is also studied in 
\cite{Gumrukcuoglu:2007bx}. These two studies have shown that, though the perturbation theory is well-defined for a scalar field in an anisotropic background, the quantisation of the canonical degrees of freedom is lacking from a well-defined asymptotic 
adiabatic vacuum. Indeed the fact that a scalar field has no anisotropic stress implies that the shear is decreasing, which means that 
it necessarily increases when going back in time. As a result the perturbation theory fails to be predictive above a certain 
scale which corresponds to the modes which have exited the Hubble radius when the anisotropy was still largely dominating. However for such models, the power spectrum converges toward a nearly scale-invariant shape for scales smaller than this limiting scale, and if inflation has lasted long enough the larger modes can still be outside the Hubble horizon in the present universe and we recover the usual predictions obtained when the background homogeneous cosmology is taken with the FRW symmetries.

\subsection{1-form}
\label{p_1}

Performing the full perturbation theory can be very tedious. Here we only try
to find the main behaviour of the result by simplifying the perturbation
scheme. We thus ignore the backreactions and ignore the perturbations in the
metric and work on a Minkowski background. Since the gauge invariant variables involve a combination of the perturbations of the field and of the metric, this is not a bad approximation. This boils down to consider a test field with a mass directly 
related to the nonminimal coupling. This is similar to what is undertaken in \cite{Himmetoglu:2008hx}. (About nongaussianity in 
of perturbations in the presence of vector fields, see \cite{Yokoyama:2008xw,dim,Dimopoulos:2008yv}). \\

We start from the action (\ref{action}) with $V(x)=\frac{1}{2}M^2x^2$ and a
Minkowski background, and (see \cite{Himmetoglu:2008hx})
setting $A_M=(\alpha_0,\partial_i \alpha + \alpha_i)$ we obtain
\bea
S &=& \int d \eta d^3 k \left\{ \frac{1}{2}\left[|\alpha_i'|^2-\left(k^2+M^2
    \right)|\alpha_i|^2 \right]+\frac{1}{2}\left[ k^2|\alpha'|^2
  \right. \right. \nonumber \\ &-& \left.\left. k^2({\alpha'}^*
    \alpha_0+cc)-M^2 k^2|\alpha|^2+\left(
    k^2+M^2\right)|\alpha_0|^2\right]\right\}\nonumber 
\eea
we obtain a constraint from $\alpha_0$ and plugging it back the action is recast as
\bea
S = \int d \eta d^3 k \frac{k^2 M^2}{2}\left[ \frac{|\alpha'|^2}{k^2+M^2}-|\alpha|^2\right]
+\int d \eta d^3 k \frac{1}{2}\left[|\alpha_i'|^2-\left(k^2+M^2\right)|\alpha_i|^2\right]
\eea
We conclude that if $M^2 < 0$ the mode $\alpha$ is not well behaved but it is if $M^2>0$. 
See however the discussion \ref{disc}.

One more way to look at the possible appearance of the ghost mode is to employ the dual description of 
the section \ref{dual}. The non-minimally coupled vector Lagrangian is dual to a massive two-form $B$ 
with a noncanonical kinetic term. The Lagrangian for this two-form $B$ reads explicitly
\be
\mathcal{L} = -\frac{3}{\xi R/m^2+1}F^2(B) - \frac{1}{12}m^2 B^2.
\ee
Thus, since now the prefactor of the kinetic term becomes negative for $\xi<0$ and $m^2 < R$ it is a ghost, as least as long as
$R$ may be regarded as a constant background field.
This also clarifies the similarities between vector field and two-form inflation. The vector inflation ${\it is}$ the two-form
inflation, where one has a bare mass term but a ghost (or in general, more complicated) kinetic term. 

\subsection{2-form}
\label{p_2}

We start from the action (\ref{action}) in Minkowski background and with
$V(x)=\frac{1}{4}x^2$. We then decompose $A_{M N}$ in the following manner 
\bea
A_{0i}=\partial_i E + E_i\qquad \qquad
A_{ij}=\epsilon_{ijk}(\partial^k B + B^k)
\eea
with $\partial_i E^i=\partial_i B^i=0$.
Now the perturbed action is, up to total derivatives

\bea
S&=&\int d\eta d^3k  \left[ \frac{1}{2} B_i' {B^i}' +\frac{1}{2}\partial_i
  B' \partial^i B'-\frac{1}{2}\Delta B \Delta B+\frac{1}{2} \partial_i
  E_j \partial^i E^j-{B^i}'\epsilon_{ijk}\partial^k E^j\right]\\
&&+\int d^4 x \left[ -\frac{1}{2}M^2 B_i B^i+\frac{1}{2}M^2 E_i
  E^i-\frac{1}{2}M^2 \partial_i B \partial^i B+\frac{1}{2}M^2 \partial_i
  E \partial^i E\right]
\eea
We can go in Fourier space and decompose the vector terms on an orthonormal basis $e_1^i, e_2^i,\hat k^i$ 
\be
B^i=\sum_{a=1,2} {\rm i}B^a  e_a^i.
\ee
As constraints, in Fourier space, we obtain
\bea
E&=&0\\
(M^2+k^2) E_a, &=& {\cal M}_a^{\,\,b} k B_b'
\eea
where 
\be
{\cal M}_a^{\,\,b}=\left(\begin{array}{cc}
             0 &-1\\1&0
             \end{array}\right).
\ee
So $E$ and $E_i$ are constrained and once replaced in the action we obtain
\bea
S=\int d\eta d^3 k \left\{ \frac{k^2}{2}\left[|B'|^2-\left(k^2+M^2
    \right)|B|^2 \right] + \frac{M^2}{2}\left[\frac{|{B_i}'|^2}{k^2+M^2} -|B_i|^2\right]\right\}\,.
\eea
So now the well behaved part is $B$ and the part which is not well behaved when $M^2<0$ is $B_i$ but the argument is similar. See the discussion in the next section about it. 
 
Again, like in the vector case, we can exploit at the dual description derived the section \ref{dual} to recast the two-form into a vector. We already know
that the non-minimal coupling will become a kinetic coupling and that the canonical kinetic term transforms into canonical mass term.
Writing then explicitly the Lagrangian for the vector $B$ which is equivalent to the two-form reads, 
\be
\mathcal{L} = -\frac{1}{\xi R/m^2+1}F^2(B) - \frac{1}{2}m^2 B^2.
\ee
In the two-form decription, one has to perform the decomposition or use the St\"uckelberg trick to fish out the vector degree of freedom that appears 
then with the wrong sign. Alternatively however, one may use the dual description as a vector field and see immediately when the ghost seems to appear. Now, 
of course again the prefactor of the kinetic term becomes negative for $\xi<0$ and $m^2 < R$. 

\subsection{Three-form}

The arguments of the section \ref{dual} and of the section \ref{gauge} both support the stability of the three-form, whereas in the vector and two-form 
cases both seemed to have problematical implications. More specifically, the gauge symmetry completion of the three-form with positive mass is a
canonical two-form. Also, the dual scalar field is not a ghost unless the mass was negative. Also, the three-form cosmology can do without anisotropy 
and nonminimal couplings, which may otherwise introduce problems with stability. This strongly supports the conclusion that three-forms models ${\it can}$
be viable. Investigation of more specific models will be presented elsewhere \cite{nelson,nelson2}.

\subsection{About the nature of instability}
\label{disc}

From our analysis in subsection (\ref{gauge}), it is clear that in flat spacetime with constant $M^2<0$ there is a ghost. However, it is more subtle to see how this analogy can be used to analyse a nonminimal coupling in FRW spacetime which is not Ricci-flat.
An argument for not using this analogy might seem to be that the gravitational
degrees of freedom in $R$ have to integrated out in order to write the
quadratic action in a canonical form. The full computation which consists in
perturbing quadratically both the metric and the field has been performed in
Ref~\cite{Himmetoglu:2008hx}, though for the specific model of
Ref~\cite{Ackerman:2007nb} (see also (\cite{boe}). What is found is that the coefficient in front of the kinetic term changes its sign. When it 
does so, the solution could be expected to diverge. The appearance of ghost seems to be linked with a classical divergence. 
This is understandable, since if there is no consistent classical solution, there is no possibility to quantise. 
Similar relation between classical singular behaviour and an appearance of a quantum ghost has been found in the 
Gauss-Bonnet cosmology \cite{Nojiri:2005vv}.
There also a divergence of cosmological perturbations (which was first noticed in the context of first-loop string cosmology 
\cite{Kawai:1998ab}) is directly linked with the change of the sign of a kinetic term of an effective degree of 
freedom \cite{Koivisto:2006ai,Koivisto:2006xf}. 
In the case of higher inverse 
derivative gravity \cite{Nojiri:2007uq}, the cosmological background reaches a sudden singularity at the moment that a sign flips in 
front of a scalar 
degree of freedom of the theory \cite{Koivisto:2008xfa,Koivisto:2008dh}. Again a classical divergence reflects a fundamental quantum 
problem.

However, the divergence of the classical solution which was expected to appear around horizon crossing has been claimed to in fact,
not be there in the solutions which were presented explicitly in Ref.~\cite{Dimopoulos:2008yv} where the longitudinal mode was also analysed in detail. It 
was also argued that the ghost might not be 
dangerous to the theory if it appears
at a time when all the couplings to other field are negligibly small.   
No divergence was directly seen in Ref~\cite{Golovnev:2009ks}. The question then arises whether the apparent classical and quantum problems both
disappear at more careful scrutiny. Finally, we remind that though the above condiserations are made for a vector field, by the strong analogy we have 
developed during this paper, they apply (almost) as such to the two-form as well.



\section{Conclusions}
\label{conclu}

In this work we have investigated anisotropic and isotropic slow-roll
inflation supported by n-forms. We have used both canonical field strengths
and their duals, allowing them to have a potential and a nonminimal coupling
to curvature if necessary. New anisotropic solutions were found for two-form,
generalising vector inflation. However, we have found that, as in the case of
vector inflation, inflation driven by the two-form would be unstable. One type
of problem is that a new field appears into the model which is pathological
when the effective mass squared is negative. We showed this also by
considering the action for perturbations in the simple flat case. In
addition, the stability of a background with initially small shear was
investigated, which for one-form field seems critical but with a triad more
reasonable.

New viable isotropic solutions were found. Three-forms could naturally support inflation. In particular, the three-form could inflate even if 
the field wasn't slowly rolling and could realise a phantom inflaton depending on the slope of the potential. 
Moreover, we have also shown that some $n$-form actions are equivalent to $f(R)$ gravity and some to scalar field models with a possible nonminimal coupling. These 
observations seem to strongly motivate further investigation of form-driven inflation. In particular, one is interested in the observational predictions for 
the nature of primordial perturbations and their non-Gaussianity.

More compactly, the two main lessons of our study are the following. 
\begin{itemize}
\item The assumption that inflaton is driven by a scalar field is not robust. We have found simple and viable inflationary models driven by a form field.
\item The assumption that the inflaton appears isotropic at large scales seems justified. The anisotropic solutions in the simplest cases seem to generically require 
nonminimal couplings and feature instabilities.
\end{itemize}
The findings in each of the five cases are summarised in Table \ref{tab}. 
On the more formal side, a reformulation of the dual forms with nonminimal coupling was proposed which seems 
to generalise some scalar-tensor models in a way which could be possibly useful in the context of alternative dark matter theories and the cosmological constant
problem. Let us remark that though the inflaton seems isotropic, anisotropy could originate from fields playing role as impurity 
during inflaton or curvaton generating the perturbations \cite{Dimopoulos:2006ms,Kanno:2008gn,Koh:2009vm}. In particular, gauge 
invariance preserving form field could naturally leave percent level anisotropic hair if nonminimally coupled with the inflaton 
\cite{Watanabe:2009ct}.

It would be interesting to apply these considerations during the dark energy
era. Then it becomes relevant 
to study the dynamics of anisotropic component in the presence of matter
fluids \cite{Koivisto:2007bp,Koivisto:2008ig} and 
its effects to the formation of large scale structure
\cite{Koivisto:2005mm,Mota:2007sz}. Indeed, possible anisotropic effects 
in the CMB might be due to an unexpected property of the late acceleration,
they do not have to be imprinted already in the 
primordial inflationary spectrum of fluctuations. Furthermore, interesting
constraints could be potentially obtained from possible CMB B-modes
polarisation within these anisotropic cosmologies. Such possibility is enhanced
from the transformation of E-modes to B-modes due to the shear \cite{Pontzen:2007ii}.

\begin{acknowledgments} 
We would like to Carsten van de Bruck for useful discussions. TSK thanks the Galileo Galilei Institute for Theoretical Physics for hospitality 
and the INFN for partial support during the completion of this work. DFM acknowledges support from the CNRS Visitor Research Fellowship at the University of 
Montpellier.
\end{acknowledgments}

\bibliography{vec}

\providecommand{\href}[2]{#2}\begingroup\raggedright\begin{thebibliography}{10}

\bibitem{inf1}
A.~H. Guth, {\it {The Inflationary Universe: A Possible Solution to the Horizon
  and Flatness Problems}},  {\em Phys. Rev.} {\bf D23} (1981) 347--356.

\bibitem{inf2}
S.~Kachru {\em et.~al.}, {\it {Towards inflation in string theory}},  {\em
  JCAP} {\bf 0310} (2003) 013,
  [\href{http://xxx.lanl.gov/abs/hep-th/0308055}{{\tt hep-th/0308055}}].

\bibitem{inf3}
D.~A. Easson, R.~Gregory, D.~F. Mota, G.~Tasinato, and I.~Zavala, {\it
  {Spinflation}},  {\em JCAP} {\bf 0802} (2008) 010,
  [\href{http://xxx.lanl.gov/abs/0709.2666}{{\tt arXiv:0709.2666}}].

\bibitem{inf4}
G.~R. Dvali and S.~H.~H. Tye, {\it {Brane inflation}},  {\em Phys. Lett.} {\bf
  B450} (1999) 72--82, [\href{http://xxx.lanl.gov/abs/hep-ph/9812483}{{\tt
  hep-ph/9812483}}].

\bibitem{inf5}
D.~H. Lyth and A.~Riotto, {\it {Particle physics models of inflation and the
  cosmological density perturbation}},  {\em Phys. Rept.} {\bf 314} (1999)
  1--146, [\href{http://xxx.lanl.gov/abs/hep-ph/9807278}{{\tt
  hep-ph/9807278}}].

\bibitem{inf7}
L.~Kofman, A.~D. Linde, and A.~A. Starobinsky, {\it {Reheating after
  inflation}},  {\em Phys. Rev. Lett.} {\bf 73} (1994) 3195--3198,
  [\href{http://xxx.lanl.gov/abs/hep-th/9405187}{{\tt hep-th/9405187}}].

\bibitem{inf8}
E.~J. Copeland, A.~R. Liddle, D.~H. Lyth, E.~D. Stewart, and D.~Wands, {\it
  {False vacuum inflation with Einstein gravity}},  {\em Phys. Rev.} {\bf D49}
  (1994) 6410--6433, [\href{http://xxx.lanl.gov/abs/astro-ph/9401011}{{\tt
  astro-ph/9401011}}].

\bibitem{inf9}
A.~D. Linde, {\it {Hybrid inflation}},  {\em Phys. Rev.} {\bf D49} (1994)
  748--754, [\href{http://xxx.lanl.gov/abs/astro-ph/9307002}{{\tt
  astro-ph/9307002}}].

\bibitem{inf10}
C.~P. Burgess, R.~Easther, A.~Mazumdar, D.~F. Mota, and T.~Multamaki, {\it
  {Multiple inflation, cosmic string networks and the string landscape}},  {\em
  JHEP} {\bf 05} (2005) 067,
  [\href{http://xxx.lanl.gov/abs/hep-th/0501125}{{\tt hep-th/0501125}}].

\bibitem{inf11}
A.~D. Linde, {\it {Chaotic Inflation}},  {\em Phys. Lett.} {\bf B129} (1983)
  177--181.

\bibitem{Koivisto:2008xf}
T.~S. Koivisto and D.~F. Mota, {\it {Vector Field Models of Inflation and Dark
  Energy}},  {\em JCAP} {\bf 0808} (2008) 021,
  [\href{http://xxx.lanl.gov/abs/0805.4229}{{\tt arXiv:0805.4229}}].

\bibitem{Golovnev:2008cf}
A.~Golovnev, V.~Mukhanov, and V.~Vanchurin, {\it {Vector Inflation}},  {\em
  JCAP} {\bf 0806} (2008) 009, [\href{http://xxx.lanl.gov/abs/0802.2068}{{\tt
  arXiv:0802.2068}}].

\bibitem{Ford:1989me}
L.~H. Ford, {\it {Inflation driven by a vector field}},  {\em Phys. Rev.} {\bf
  D40} (1989) 967.

\bibitem{Turner:1987bw}
M.~S. Turner and L.~M. Widrow, {\it {Inflation Produced, Large Scale Magnetic
  Fields}},  {\em Phys. Rev.} {\bf D37} (1988) 2743.

\bibitem{bertolami}
O. Bertolami and D. F. Mota, {\it {Primordial Magnetic Fields via Spontaneous Breaking of Lorentz Invariance}},  {\em Phys. Lett.} {\bf B455} (1998) 96.

\bibitem{c4}
H.~K. Eriksen, F.~K. Hansen, A.~J. Banday, K.~M. Gorski, and P.~B. Lilje, {\it
  {Asymmetries in the CMB anisotropy field}},  {\em Astrophys. J.} {\bf 605}
  (2004) 14--20, [\href{http://xxx.lanl.gov/abs/astro-ph/0307507}{{\tt
  astro-ph/0307507}}].

\bibitem{c3}
{\bf WMAP} Collaboration, G.~Hinshaw {\em et.~al.}, {\it {Three-year Wilkinson
  Microwave Anisotropy Probe (WMAP) observations: Temperature analysis}},  {\em
  Astrophys. J. Suppl.} {\bf 170} (2007) 288,
  [\href{http://xxx.lanl.gov/abs/astro-ph/0603451}{{\tt astro-ph/0603451}}].

\bibitem{c2}
K.~Land and J.~Magueijo, {\it {The axis of evil}},  {\em Phys. Rev. Lett.} {\bf
  95} (2005) 071301, [\href{http://xxx.lanl.gov/abs/astro-ph/0502237}{{\tt
  astro-ph/0502237}}].

\bibitem{c1}
A.~Rakic and D.~J. Schwarz, {\it {Correlating anomalies of the microwave sky:
  The Good, the Evil and the Axis}},  {\em Phys. Rev.} {\bf D75} (2007) 103002,
  [\href{http://xxx.lanl.gov/abs/astro-ph/0703266}{{\tt astro-ph/0703266}}].

\bibitem{f1}
G.~Efstathiou, {\it {A Maximum Likelihood Analysis of the Low CMB Multipoles
  from WMAP}},  {\em Mon. Not. Roy. Astron. Soc.} {\bf 348} (2004) 885,
  [\href{http://xxx.lanl.gov/abs/astro-ph/0310207}{{\tt astro-ph/0310207}}].

\bibitem{f2}
A.~de~Oliveira-Costa, M.~Tegmark, M.~Zaldarriaga, and A.~Hamilton, {\it {The
  significance of the largest scale CMB fluctuations in WMAP}},  {\em Phys.
  Rev.} {\bf D69} (2004) 063516,
  [\href{http://xxx.lanl.gov/abs/astro-ph/0307282}{{\tt astro-ph/0307282}}].

\bibitem{f3}
J.~Magueijo and R.~D. Sorkin, {\it {Occam's razor meets WMAP}},  {\em Mon. Not.
  Roy. Astron. Soc. Lett.} {\bf 377} (2007) L39--L43,
  [\href{http://xxx.lanl.gov/abs/astro-ph/0604410}{{\tt astro-ph/0604410}}].

\bibitem{f4}
P.~Bielewicz, K.~M. Gorski, and A.~J. Banday, {\it {Low order multipole maps of
  CMB anisotropy derived from WMAP}},  {\em Mon. Not. Roy. Astron. Soc.} {\bf
  355} (2004) 1283, [\href{http://xxx.lanl.gov/abs/astro-ph/0405007}{{\tt
  astro-ph/0405007}}].

\bibitem{frode1}
J.~Hoftuft {\em et.~al.}, {\it {Increasing evidence for hemispherical power
  asymmetry in the five-year WMAP data}},
  \href{http://xxx.lanl.gov/abs/0903.1229}{{\tt arXiv:0903.1229}}.

\bibitem{frode2}
F.~K. Hansen, A.~J. Banday, K.~M. Gorski, H.~K. Eriksen, and P.~B. Lilje, {\it
  {Power Asymmetry in Cosmic Microwave Background Fluctuations from Full Sky to
  Sub-degree Scales: Is the Universe Isotropic?}},
  \href{http://xxx.lanl.gov/abs/0812.3795}{{\tt arXiv:0812.3795}}.

\bibitem{nic}
N.~E. Groeneboom and H.~K. Eriksen, {\it {Bayesian analysis of sparse
  anisotropic universe models and application to the 5-yr WMAP data}},  {\em
  Astrophys. J.} {\bf 690} (2009) 1807--1819,
  [\href{http://xxx.lanl.gov/abs/0807.2242}{{\tt arXiv:0807.2242}}].

\bibitem{s1}
S.~L. Bridle, A.~M. Lewis, J.~Weller, and G.~Efstathiou, {\it {Reconstructing
  the primordial power spectrum}},  {\em Mon. Not. Roy. Astron. Soc.} {\bf 342}
  (2003) L72, [\href{http://xxx.lanl.gov/abs/astro-ph/0302306}{{\tt
  astro-ph/0302306}}].

\bibitem{s2}
D.~J. Schwarz, G.~D. Starkman, D.~Huterer, and C.~J. Copi, {\it {Is the low-l
  microwave background cosmic?}},  {\em Phys. Rev. Lett.} {\bf 93} (2004)
  221301, [\href{http://xxx.lanl.gov/abs/astro-ph/0403353}{{\tt
  astro-ph/0403353}}].

\bibitem{s3}
A.~Slosar, U.~Seljak, and A.~Makarov, {\it {Exact likelihood evaluations and
  foreground marginalization in low resolution WMAP data}},  {\em Phys. Rev.}
  {\bf D69} (2004) 123003,
  [\href{http://xxx.lanl.gov/abs/astro-ph/0403073}{{\tt astro-ph/0403073}}].

\bibitem{s4}
S.~Prunet, J.-P. Uzan, F.~Bernardeau, and T.~Brunier, {\it {Constraints on mode
  couplings and modulation of the CMB with WMAP data}},  {\em Phys. Rev.} {\bf
  D71} (2005) 083508, [\href{http://xxx.lanl.gov/abs/astro-ph/0406364}{{\tt
  astro-ph/0406364}}].

\bibitem{Germani:2009iq}
C.~Germani and A.~Kehagias, {\it {P-nflation: generating cosmic Inflation with
  p-forms}},  \href{http://xxx.lanl.gov/abs/0902.3667}{{\tt arXiv:0902.3667}}.

\bibitem{Kobayashi:2009hj}
T.~Kobayashi and S.~Yokoyama, {\it {Gravitational waves from p-form
  inflation}},  \href{http://xxx.lanl.gov/abs/0903.2769}{{\tt
  arXiv:0903.2769}}.

\bibitem{germani2} C. Germani and A. Kehagias, {\it Scalar perturbations in p-nflation: the 3-form case}, \href{http://xxx.lanl.gov/abs/0908.0001}{{\tt arXiv:0908.0001}}.
 
\bibitem{dimopolous} K. Dimopoulos, M. KarciauskasandJ. M. Wagstaﬀ, {\it  Vector Curvaton with varying Kinetic Function}, \href{http://xxx.lanl.gov/abs/0907.1838}{{\tt arXiv:0907.1838}}. 

\bibitem{Moffat:1994hv}
J.~W. Moffat, {\it {Nonsymmetric gravitational theory}},  {\em Phys. Lett.}
  {\bf B355} (1995) 447--452,
  [\href{http://xxx.lanl.gov/abs/gr-qc/9411006}{{\tt gr-qc/9411006}}].

\bibitem{Prokopec:2006kr}
T.~Prokopec and W.~Valkenburg, {\it {Antisymmetric metric field as dark
  matter}},  \href{http://xxx.lanl.gov/abs/astro-ph/0606315}{{\tt
  astro-ph/0606315}}.

\bibitem{Damour:1992bt}
T.~Damour, S.~Deser, and J.~G. McCarthy, {\it {Nonsymmetric gravity theories:
  Inconsistencies and a cure}},  {\em Phys. Rev.} {\bf D47} (1993) 1541--1556,
  [\href{http://xxx.lanl.gov/abs/gr-qc/9207003}{{\tt gr-qc/9207003}}].

\bibitem{Clayton:1996dz}
M.~A. Clayton, {\it {Linearisation Instabilities of the Massive Nonsymmetric
  Gravitational Theory}},  {\em Class. Quant. Grav.} {\bf 13} (1996)
  2851--2864, [\href{http://xxx.lanl.gov/abs/gr-qc/9603062}{{\tt
  gr-qc/9603062}}].

\bibitem{Prokopec:2005fb}
T.~Prokopec and W.~Valkenburg, {\it {The cosmology of the nonsymmetric theory
  of gravitation}},  {\em Phys. Lett.} {\bf B636} (2006) 1--4,
  [\href{http://xxx.lanl.gov/abs/astro-ph/0503289}{{\tt astro-ph/0503289}}].

\bibitem{ArmendarizPicon:2003qw}
C.~Armendariz-Picon and V.~Duvvuri, {\it {Anisotropic inflation and the origin
  of four large dimensions}},  {\em Class. Quant. Grav.} {\bf 21} (2004)
  2011--2028, [\href{http://xxx.lanl.gov/abs/hep-th/0305237}{{\tt
  hep-th/0305237}}].

\bibitem{VanNieuwenhuizen:1973fi}
P.~Van~Nieuwenhuizen, {\it {On ghost-free tensor lagrangians and linearized
  gravitation}},  {\em Nucl. Phys.} {\bf B60} (1973) 478--492.

\bibitem{Drummond:1979pp}
I.~T. Drummond and S.~J. Hathrell, {\it {QED Vacuum Polarization in a
  Background Gravitational Field and Its Effect on the Velocity of Photons}},
  {\em Phys. Rev.} {\bf D22} (1980) 343.

\bibitem{Bamba:2008xa}
K.~Bamba, S.~Nojiri, and S.~D. Odintsov, {\it {Inflationary cosmology and the
  late-time accelerated expansion of the universe in non-minimal
  Yang-Mills-$F(R)$ gravity and non-minimal vector-$F(R)$ gravity}},  {\em
  Phys. Rev.} {\bf D77} (2008) 123532,
  [\href{http://xxx.lanl.gov/abs/0803.3384}{{\tt arXiv:0803.3384}}].

\bibitem{Bamba:2008ja}
K.~Bamba and S.~D. Odintsov, {\it {Inflation and late-time cosmic acceleration
  in non-minimal Maxwell-$F(R)$ gravity and the generation of large-scale
  magnetic fields}},  {\em JCAP} {\bf 0804} (2008) 024,
  [\href{http://xxx.lanl.gov/abs/0801.0954}{{\tt arXiv:0801.0954}}].

\bibitem{Janssen:2006nn}
T.~Janssen and T.~Prokopec, {\it {Instabilities in the nonsymmetric theory of
  gravitation}},  {\em Class. Quant. Grav.} {\bf 23} (2006) 4967--4982,
  [\href{http://xxx.lanl.gov/abs/gr-qc/0604094}{{\tt gr-qc/0604094}}].

\bibitem{Eguchi:1980jx}
T.~Eguchi, P.~B. Gilkey, and A.~J. Hanson, {\it {Gravitation, Gauge Theories
  and Differential Geometry}},  {\em Phys. Rept.} {\bf 66} (1980) 213.

\bibitem{ArmendarizPicon:1999rj}
C.~Armendariz-Picon, T.~Damour, and V.~F. Mukhanov, {\it {k-Inflation}},  {\em
  Phys. Lett.} {\bf B458} (1999) 209--218,
  [\href{http://xxx.lanl.gov/abs/hep-th/9904075}{{\tt hep-th/9904075}}].

\bibitem{Gruzinov:2004rq}
A.~Gruzinov, {\it {Three form inflation}},  {\em astro-ph/0401520} (2004).

\bibitem{Dolgov:2003fw}
A.~D. Dolgov and M.~Kawasaki, {\it {Realistic cosmological model with dynamical
  cancellation of vacuum energy}},
  \href{http://xxx.lanl.gov/abs/astro-ph/0307442}{{\tt astro-ph/0307442}}.

\bibitem{Nojiri:2004bi}
S.~Nojiri and S.~D. Odintsov, {\it {Gravity assisted dark energy dominance and
  cosmic acceleration}},  {\em Phys. Lett.} {\bf B599} (2004) 137--142,
  [\href{http://xxx.lanl.gov/abs/astro-ph/0403622}{{\tt astro-ph/0403622}}].

\bibitem{Koivisto:2005yk}
T.~Koivisto, {\it {Covariant conservation of energy momentum in modified
  gravities}},  {\em Class. Quant. Grav.} {\bf 23} (2006) 4289--4296,
  [\href{http://xxx.lanl.gov/abs/gr-qc/0505128}{{\tt gr-qc/0505128}}].

\bibitem{Sotiriou:2008it}
T.~P. Sotiriou and V.~Faraoni, {\it {Modified gravity with R-matter couplings
  and (non- )geodesic motion}},  {\em Class. Quant. Grav.} {\bf 25} (2008)
  205002, [\href{http://xxx.lanl.gov/abs/0805.1249}{{\tt arXiv:0805.1249}}].

\bibitem{Brans:1961sx}
C.~Brans and R.~H. Dicke, {\it {Mach's principle and a relativistic theory of
  gravitation}},  {\em Phys. Rev.} {\bf 124} (1961) 925--935.

\bibitem{Wetterich:1987fm}
C.~Wetterich, {\it {Cosmology and the Fate of Dilatation Symmetry}},  {\em
  Nucl. Phys.} {\bf B302} (1988) 668.

\bibitem{muk}
V.~F. Mukhanov, H.~A. Feldman, and R.~H. Brandenberger, {\it {Theory of
  cosmological perturbations. Part 1. Classical perturbations. Part 2. Quantum
  theory of perturbations. Part 3. Extensions}},  {\em Phys. Rept.} {\bf 215}
  (1992) 203--333.

\bibitem{cop}
E.~J. Copeland, M.~Sami, and S.~Tsujikawa, {\it {Dynamics of dark energy}},
  {\em Int. J. Mod. Phys.} {\bf D15} (2006) 1753--1936,
  [\href{http://xxx.lanl.gov/abs/hep-th/0603057}{{\tt hep-th/0603057}}].



\bibitem{nelson} Tomi S. Koivisto, Nelson J. Nunes, {\it Three-form cosmology.}
\href{http://xxx.lanl.gov/abs/0907.3883}{{\tt arXiv:0907.3883}}


\bibitem{nelson2} Tomi S.Koivisto Nelson J. Nunes, {\it Inflation and dark energy from three-forms.}, \href{http://xxx.lanl.gov/abs/0908.0920}
{{\tt arXiv:0908.0920}}


\bibitem{Himmetoglu:2008hx}
B.~Himmetoglu, C.~R. Contaldi, and M.~Peloso, {\it {Instability of the ACW
  model, and problems with massive vectors during inflation}},
  \href{http://xxx.lanl.gov/abs/0812.1231}{{\tt arXiv:0812.1231}}.

\bibitem{cus}
N.~Afshordi, D.~J.~H. Chung, M.~Doran, and G.~Geshnizjani, {\it {Cuscuton
  Cosmology: Dark Energy meets Modified Gravity}},  {\em Phys. Rev.} {\bf D75}
  (2007) 123509, [\href{http://xxx.lanl.gov/abs/astro-ph/0702002}{{\tt
  astro-ph/0702002}}].

\bibitem{cus2}
G.~Robbers, N.~Afshordi, and M.~Doran, {\it {Does Planck mass run on the
  cosmological horizon scale?}},  {\em Phys. Rev. Lett.} {\bf 100} (2008)
  111101, [\href{http://xxx.lanl.gov/abs/0708.3235}{{\tt arXiv:0708.3235}}].

\bibitem{Koivisto:2005yc}
T.~Koivisto and H.~Kurki-Suonio, {\it {Cosmological perturbations in the
  Palatini formulation of modified gravity}},  {\em Class. Quant. Grav.} {\bf
  23} (2006) 2355--2369, [\href{http://xxx.lanl.gov/abs/astro-ph/0509422}{{\tt
  astro-ph/0509422}}].

\bibitem{Koivisto:2007sq}
T.~Koivisto, {\it {Viable Palatini-f(R) cosmologies with generalized dark
  matter}},  {\em Phys. Rev.} {\bf D76} (2007) 043527,
  [\href{http://xxx.lanl.gov/abs/0706.0974}{{\tt arXiv:0706.0974}}].

\bibitem{pal1}
B.~Li, D.~F. Mota, and D.~J. Shaw, {\it {Microscopic and Macroscopic Behaviors
  of Palatini Modified Gravity Theories}},  {\em Phys. Rev.} {\bf D78} (2008)
  064018, [\href{http://xxx.lanl.gov/abs/0805.3428}{{\tt arXiv:0805.3428}}].

\bibitem{pal2}
B.~Li, D.~F. Mota, and D.~J. Shaw, {\it {Indistinguishable Macroscopic
  Behaviour of Palatini Gravities and General Relativity}},  {\em Class. Quant.
  Grav.} {\bf 26} (2009) 055018, [\href{http://xxx.lanl.gov/abs/0801.0603}{{\tt
  arXiv:0801.0603}}].

\bibitem{Turok:1998he}
N.~Turok and S.~W. Hawking, {\it {Open inflation, the four form and the
  cosmological constant}},  {\em Phys. Lett.} {\bf B432} (1998) 271--278,
  [\href{http://xxx.lanl.gov/abs/hep-th/9803156}{{\tt hep-th/9803156}}].

\bibitem{Pereira:2007yy}
T.~S. Pereira, C.~Pitrou, and J.-P. Uzan, {\it {Theory of cosmological
  perturbations in an anisotropic universe}},  {\em JCAP} {\bf 0709} (2007)
  006, [\href{http://xxx.lanl.gov/abs/0707.0736}{{\tt arXiv:0707.0736}}].

\bibitem{Pitrou:2008gk}
C.~Pitrou, T.~S. Pereira, and J.-P. Uzan, {\it {Predictions from an anisotropic
  inflationary era}},  {\em JCAP} {\bf 0804} (2008) 004,
  [\href{http://xxx.lanl.gov/abs/0801.3596}{{\tt arXiv:0801.3596}}].

\bibitem{Gumrukcuoglu:2007bx}
A.~E. Gumrukcuoglu, C.~R. Contaldi, and M.~Peloso, {\it {Inflationary
  perturbations in anisotropic backgrounds and their imprint on the CMB}},
  {\em JCAP} {\bf 0711} (2007) 005,
  [\href{http://xxx.lanl.gov/abs/0707.4179}{{\tt arXiv:0707.4179}}].

\bibitem{Yokoyama:2008xw}
S.~Yokoyama and J.~Soda, {\it {Primordial statistical anisotropy generated at
  the end of inflation}},  {\em JCAP} {\bf 0808} (2008) 005,
  [\href{http://xxx.lanl.gov/abs/0805.4265}{{\tt arXiv:0805.4265}}].

\bibitem{dim}
K.~Dimopoulos and M.~Karciauskas, {\it {Non-minimally coupled vector
  curvaton}},  {\em JHEP} {\bf 07} (2008) 119,
  [\href{http://xxx.lanl.gov/abs/0803.3041}{{\tt arXiv:0803.3041}}].

\bibitem{Dimopoulos:2008yv}
K.~Dimopoulos, D.~H. Lyth, M.~Karciauskas and Y.~Rodriguez, {\it {Statistical anisotropy of
  the curvature perturbation from vector field perturbations}},{\em JCAP} {\bf 0905} (2009) 013,
  \href{http://xxx.lanl.gov/abs/0809.1055}{{\tt arXiv:0809.1055}}.

\bibitem{Linde:1981mu}
A.~D. Linde, {\it {A New Inflationary Universe Scenario: A Possible Solution of
  the Horizon, Flatness, Homogeneity, Isotropy and Primordial Monopole
  Problems}},  {\em Phys. Lett.} {\bf B108} (1982) 389--393.

\bibitem{Starobinsky:1980te}
A.~A. Starobinsky, {\it {A new type of isotropic cosmological models without
  singularity}},  {\em Phys. Lett.} {\bf B91} (1980) 99--102.

\bibitem{Ackerman:2007nb}
L.~Ackerman, S.~M. Carroll, and M.~B. Wise, {\it {Imprints of a Primordial
  Preferred Direction on the Microwave Background}},  {\em Phys. Rev.} {\bf
  D75} (2007) 083502, [\href{http://xxx.lanl.gov/abs/astro-ph/0701357}{{\tt
  astro-ph/0701357}}].

\bibitem{boe}
C.~G. Boehmer and D.~F. Mota, {\it {CMB Anisotropies and Inflation from
  Non-Standard Spinors}},  {\em Phys. Lett.} {\bf B663} (2008) 168--171,
  [\href{http://xxx.lanl.gov/abs/0710.2003}{{\tt arXiv:0710.2003}}].

\bibitem{Golovnev:2009ks}
A.~Golovnev and V.~Vanchurin, {\it {Cosmological perturbations from vector
  inflation}},  \href{http://xxx.lanl.gov/abs/0903.2977}{{\tt
  arXiv:0903.2977}}.

\bibitem{Nojiri:2005vv}
S.~Nojiri, S.~D. Odintsov, and M.~Sasaki, {\it {Gauss-Bonnet dark energy}},
  {\em Phys. Rev.} {\bf D71} (2005) 123509,
  [\href{http://xxx.lanl.gov/abs/hep-th/0504052}{{\tt hep-th/0504052}}].

\bibitem{Kawai:1998ab}
S.~Kawai, M.-a. Sakagami, and J.~Soda, {\it {Instability of 1-loop superstring
  cosmology}},  {\em Phys. Lett.} {\bf B437} (1998) 284--290,
  [\href{http://xxx.lanl.gov/abs/gr-qc/9802033}{{\tt gr-qc/9802033}}].

\bibitem{Koivisto:2006ai}
T.~Koivisto and D.~F. Mota, {\it {Gauss-Bonnet quintessence: Background
  evolution, large scale structure and cosmological constraints}},  {\em Phys.
  Rev.} {\bf D75} (2007) 023518,
  [\href{http://xxx.lanl.gov/abs/hep-th/0609155}{{\tt hep-th/0609155}}].

\bibitem{Koivisto:2006xf}
T.~Koivisto and D.~F. Mota, {\it {Cosmology and astrophysical constraints of
  Gauss-Bonnet dark energy}},  {\em Phys. Lett.} {\bf B644} (2007) 104--108,
  [\href{http://xxx.lanl.gov/abs/astro-ph/0606078}{{\tt astro-ph/0606078}}].

\bibitem{Nojiri:2007uq}
S.~Nojiri and S.~D. Odintsov, {\it {Modified non-local-F(R) gravity as the key
  for the inflation and dark energy}},  {\em Phys. Lett.} {\bf B659} (2008)
  821--826, [\href{http://xxx.lanl.gov/abs/0708.0924}{{\tt arXiv:0708.0924}}].

\bibitem{Koivisto:2008xfa}
T.~Koivisto, {\it {Dynamics of Nonlocal Cosmology}},  {\em Phys. Rev.} {\bf
  D77} (2008) 123513, [\href{http://xxx.lanl.gov/abs/0803.3399}{{\tt
  arXiv:0803.3399}}].

\bibitem{Koivisto:2008dh}
T.~S. Koivisto, {\it {Newtonian limit of nonlocal cosmology}},  {\em Phys.
  Rev.} {\bf D78} (2008) 123505, [\href{http://xxx.lanl.gov/abs/0807.3778}{{\tt
  arXiv:0807.3778}}].

\bibitem{Dimopoulos:2006ms}
K.~Dimopoulos, {\it {Can a vector field be responsible for the curvature
  perturbation in the universe?}},  {\em Phys. Rev.} {\bf D74} (2006) 083502,
  [\href{http://xxx.lanl.gov/abs/hep-ph/0607229}{{\tt hep-ph/0607229}}].

\bibitem{Kanno:2008gn}
S.~Kanno, M.~Kimura, J.~Soda, and S.~Yokoyama, {\it {Anisotropic Inflation from
  Vector Impurity}},  {\em JCAP} {\bf 0808} (2008) 034,
  [\href{http://xxx.lanl.gov/abs/0806.2422}{{\tt arXiv:0806.2422}}].

\bibitem{Koh:2009vm}
S.~Koh and B.~Hu, {\it {Timelike Vector Field Dynamics in the Early Universe}},
   \href{http://xxx.lanl.gov/abs/0901.0429}{{\tt arXiv:0901.0429}}.

\bibitem{Watanabe:2009ct}
M.-a. Watanabe, S.~Kanno, and J.~Soda, {\it {Hairy Inflation}},
  \href{http://xxx.lanl.gov/abs/0902.2833}{{\tt arXiv:0902.2833}}.

\bibitem{Koivisto:2007bp}
T.~Koivisto and D.~F. Mota, {\it {Accelerating Cosmologies with an Anisotropic
  Equation of State}},  {\em Astrophys. J.} {\bf 679} (2008) 1,
  [\href{http://xxx.lanl.gov/abs/0707.0279}{{\tt arXiv:0707.0279}}].

\bibitem{Koivisto:2008ig}
T.~Koivisto and D.~F. Mota, {\it {Anisotropic Dark Energy: Dynamics of
  Background and Perturbations}},  {\em JCAP} {\bf 0806} (2008) 018,
  [\href{http://xxx.lanl.gov/abs/0801.3676}{{\tt arXiv:0801.3676}}].

\bibitem{Koivisto:2005mm}
T.~Koivisto and D.~F. Mota, {\it {Dark energy anisotropic stress and large
  scale structure formation}},  {\em Phys. Rev.} {\bf D73} (2006) 083502,
  [\href{http://xxx.lanl.gov/abs/astro-ph/0512135}{{\tt astro-ph/0512135}}].

\bibitem{Mota:2007sz}
D.~F. Mota, J.~R. Kristiansen, T.~Koivisto, and N.~E. Groeneboom, {\it
  {Constraining Dark Energy Anisotropic Stress}},  {\em Mon. Not. Roy. Astron.
  Soc.} {\bf 382} (2007) 793--800,
  [\href{http://xxx.lanl.gov/abs/0708.0830}{{\tt arXiv:0708.0830}}].

\bibitem{Pontzen:2007ii}
A.~Pontzen and A.~Challinor, {\it {Bianchi Model CMB Polarization and its
  Implications for CMB Anomalies}},
  \href{http://xxx.lanl.gov/abs/0706.2075}{{\tt arXiv:0706.2075}}.

\end{thebibliography}\endgroup

\end{document}